\documentclass[prd, twocolumn, nofootinbib, floatfix]{revtex4}

\usepackage{epsfig}
\usepackage{amsmath}
\usepackage{color}

\newcommand{\beq}{\begin{equation}}
\newcommand{\eeq}{\end{equation}}
\newcommand{\beqa}{\begin{eqnarray}}
\newcommand{\eeqa}{\end{eqnarray}}

\newcommand{\lam}{\lambda}
\newcommand{\lamh}{\hat\lambda}
\newcommand{\gam}{\gamma}

\newcommand{\ls}{\mathrel{\raise0.27ex\hbox{$<$}\kern-0.70em \lower0.71ex\hbox{{
$\scriptstyle \sim$}}}}

\begin{document}

\title{Dark Energy Properties in DBI Theory}
\author{Changrim Ahn${}^{1}$, Chanju Kim${}^{1}$, and Eric V.\ Linder${}^{1,2}$}
\affiliation{${}^1$ Institute for the Early Universe and Department 
of Physics, Ewha Womans University,
Seoul 120-750, South Korea\\
${}^2$ Berkeley Center for Cosmological Physics and Berkeley Lab, 
University of California, Berkeley 94720, USA}
\date{\today}

\begin{abstract}
The Dirac-Born-Infeld (DBI) action from string theory provides 
several new classes of dark energy behavior beyond quintessence 
due to its relativistic kinematics.  We constrain parameters of 
natural potentials and brane tensions with cosmological observations 
as well as showing how to design these functions for a desired 
expansion history.  We enlarge the attractor solutions, including 
new ways of obtaining cosmological constant behavior, to the case 
of generalized DBI theory with multiple branes.  An interesting 
novel signature of DBI attractors is that the sound speed is driven 
to zero, unlike for quintessence where it is the speed of light. 
\end{abstract}

\maketitle

\section{Introduction \label{sec:intro}}

High energy physics theories for dark energy causing the accelerated 
expansion of the universe face issues of naturalness -- why is the 
current dark energy density measured so different from the initial 
conditions of the high energy, early universe, and how is the current 
low energy form of the potential energy related to the initial 
high energy form that should receive quantum corrections? 

The cosmological constant in particular suffers both problems.  Making 
the field dynamical helps.  To more fully solve the amplitude problem 
one would like an attractor solution, where the present behavior is 
largely insensitive to the exact initial conditions.  To ameliorate the 
form problem one would like a symmetry or geometric quantity that 
protects the potential, or ideally have it predicted from a fundamental 
theory such as string theory.  Quintessence models cannot achieve both 
properties, and even the attractor solutions have difficulty in naturally 
reaching a dark energy equation of state $w\approx-1$ \cite{zws} 
as indicated by cosmological observations. 

Paper 1 \cite{akl}, following the pioneering paper of \cite{martinyam}, 
highlighted the DBI class from string theory as possessing 
desirable properties to serve as dark energy.  In particular, it found 
not only the attractor solutions accessible to quintessence, but three 
new classes that could achieve or approach $w=-1$, the cosmological 
constant state.  
String theory can impose a specific non-trivial kinetic
behavior through the Dirac-Born-Infeld (DBI) action that arises naturally
in consideration of D3-brane motion within a warped compactification.
The field properties are related to the geometric 
position of a three dimensional brane within higher dimensions, and 
the brane tension and potential functions are (in principle) given 
by string theory, in particular through the AdS/CFT correspondence. 

In this paper we extend the attractor solutions as well as more fully 
considering the entire evolution and its observational consequences. 
In \S\ref{sec:quad} we examine in detail the case motivated by the 
simplest physics and find the viable regions of parameter space 
constrained through cosmological 
observations.  We show in \S\ref{sec:exphis} 
how to construct the required potential for a given 
cosmic expansion history or equation of state.  Generalizing DBI theory 
to multiple branes or non-standard branes adds a degree of freedom 
which we analyze in \S\ref{sec:wdbi}.  We explore a new window on 
constraining DBI dark energy with observations in terms of the dark 
energy sound speed  -- this gives a distinct prediction from 
quintessence -- and its effects on the matter density power spectrum 
in \S\ref{sec:sound}.

\section{Constraints on a Natural DBI Model \label{sec:quad}} 

The DBI action arises in Type IIB string theory in terms of the 
volume swept out by a D3-brane in a warped geometry, coupled to 
gravity.  The form is 
\beq 
S=\int d^4x\,\sqrt{-g}\,\left[-T(\phi)\sqrt{1-\dot\phi^2/T(\phi)}
+T(\phi)-V(\phi)\right], \label{eq:lag}
\eeq 
where we ignore the spatial derivatives of $\phi$.  
$T$ is the warped brane tension and $V$ is the potential arising 
from interactions with Ramond-Ramond fluxes or other sectors.  
See, e.g., \cite{siltong} for more details.  The kinetic factor is 
often written in terms of a Lorentz boost factor 
\beq 
\gamma\equiv\frac{1}{\sqrt{1-\dot\phi^2/T}}\,, \label{eq:gamdef} 
\eeq 
and the DBI dark energy equation of state is 
\beq 
w\equiv\frac{p_\phi}{\rho_\phi}=-\frac{\gam^{-1}-1+v}{\gam-1+v} 
\,, \label{eq:wdef} 
\eeq 
where $v=V/T$.  The nonrelativistic limit $\gamma-1\ll1$ leads 
to the quintessence action and equation of state. 

In \cite{akl} the main consideration was the critical points of the 
equations of motion and the asymptotic attractor behavior.  In this 
section we consider perhaps the most natural forms for the tension and 
potential and follow the specific dynamics throughout the history of 
the universe.  A complete string theory would predict the functions 
$T$ and $V$; while this is not available one can use known behaviors 
for certain circumstances.  For a pure AdS$_5$ geometry with radius 
$R$, the warped tension is given by 
\beq
T(\phi)=\tau\,\phi^4\,, \label{eq:tphi}
\eeq
with $\tau = 1/(g_s\tilde\lambda)$ where $g_s$ is the string coupling, 
$\alpha'$ is the inverse string tension, and $\tilde\lambda=R^4/\alpha'^2$ 
which is identified as the 't Hooft coupling in AdS/CFT correspondence. 

The potential is expected to have quadratic terms arising from the 
breaking of conformal invariance due to couplings to gravity and other 
sectors.  In addition, quartic terms enter from such interactions, 
while higher order terms are suppressed, e.g.\ by powers of $1/R$ 
\cite{siltong,ast}.  We therefore take an ansatz 
\beq 
V(\phi)=m^2\phi^2+cT=m^2\phi^2+c\tau\phi^4\,. \label{eq:vphi} 
\eeq 
Note that we take the potential to have a true zero minimum so that 
there is no intrinsic cosmological constant. 

For reference, we briefly review the equation of motion.  The 
DBI version of the Klein-Gordon equation is 
\beq 
\ddot\phi+3\gamma^{-2} H\dot\phi+\gamma^{-3}V_{,\phi}+\frac{1}{2}(3\gamma^{-2}
-2\gamma^{-3}-1)\,T_{,\phi}=0\,, 
\eeq 
where $H$ is the Hubble parameter, $V_{,\phi}=dV/d\phi$ and 
$T_{,\phi}=dT/d\phi$.  The energy-momentum tensor has perfect fluid form 
with energy density $\rho_\phi$ and pressure $p_\phi$ given by 
\beq 
\rho_\phi=(\gamma-1)\,T+V\quad ; \quad p_\phi=(1-\gamma^{-1})\,T-V\,, 
\eeq 
and so the equation of motion can also be viewed in terms of the 
continuity equation 
\beq 
\dot\rho_\phi=-3H(\rho_\phi+p_\phi)=-3H(\gamma-\gamma^{-1})\,T\,. 
\eeq 

For the form of Eq.~(\ref{eq:vphi}), the potential for large $\phi$ is 
dominated by the quartic term 
while for small $\phi$ it looks like a quadratic potential.  
\cite{akl} identified the ratio $V/T$ as particularly important for 
determining the attractor, if any.  With Eq.~(\ref{eq:tphi}) this 
implies that 
\beq 
v\equiv\frac{V}{T}=c+\mu^2\,(\kappa\phi)^{-2}\, \label{eq:vmu} 
\eeq 
where $\mu^2=(m^2\kappa^2/\tau)$ and $\kappa^2=8\pi G$. 
At late times, $\phi$ rolls to zero and the quantity $v$ is dominated 
by the second term in Eq.~(\ref{eq:vmu}) so  $v\to\infty$, giving the 
ultrarelativistic class of attractor solutions discussed by \cite{akl}. 
In particular, since $\lambda\equiv-(1/\kappa V)dV/d\phi\sim 1/\phi$ 
and $\gamma\sim v\sim\phi^{-2}$ in this limit, then the secondary 
attractor parameter of \cite{akl} is $\lam^2/\gam=$ const.  This implies 
that it is the second class of attractor solution from Table I of 
\cite{akl} that is reached and at late times $w=-1+\lam^2/(3\gam)$. 
However the evolution at present and at all times before the asymptotic 
future is of interest. 

\begin{figure}[!htb]
\begin{center}
\psfig{file=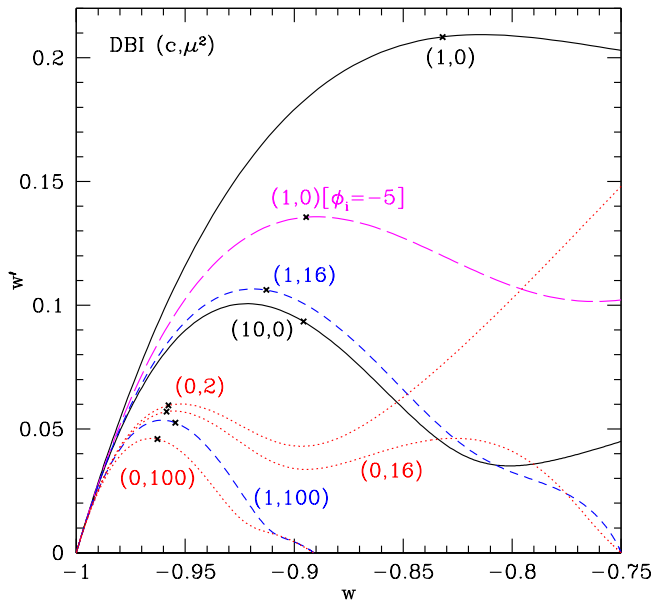,width=3.4in}
\caption{The DBI solutions using quartic/quadratic potential/tension 
functions of Eq.~(\ref{eq:vmu}) are plotted in the $w$-$w'$ plane. 
The initial thawing behavior, the values today (denoted by x's along 
the curves) with property $w'\approx 1+w$, and future attractors 
to a constant $w$ determined by the value of $\mu^2$ are all evident. 
}
\label{fig:cdmodels}
\end{center}
\end{figure}

Figure~\ref{fig:cdmodels} illustrates the dynamical evolution of 
these models in the $w$-$w'$ plane, where $w'=dw/d\ln a$, for 
various values of $c$ and $\mu^2$. 
The most noticeable common characteristic of the field evolution is 
that it is a thawing field.  That is, the dynamical history lies 
within the thawing region of the $w$-$w'$ phase 
plane defined originally for quintessence as bounded by 
$1\le w'/(1+w)\le3$, as one of the two major classes 
of evolution \cite{caldwelllinder}.  Indeed, the field evolves away 
from a frozen, $w=-1$ state in the high redshift, matter dominated 
era along the $w'=3(1+w)$ line defined by \cite{caldwelllinder} and 
shown to be a generic dynamical flow solution by \cite{cahndl}. 
The evolution remains within 
the thawing region, until today (defined by $\Omega_\phi=0.72$ and 
denoted by an x along the evolutionary track) the field lies 
roughly near $w'\approx (1+w)$.

\begin{figure}[!htb]
\begin{center}
\psfig{file=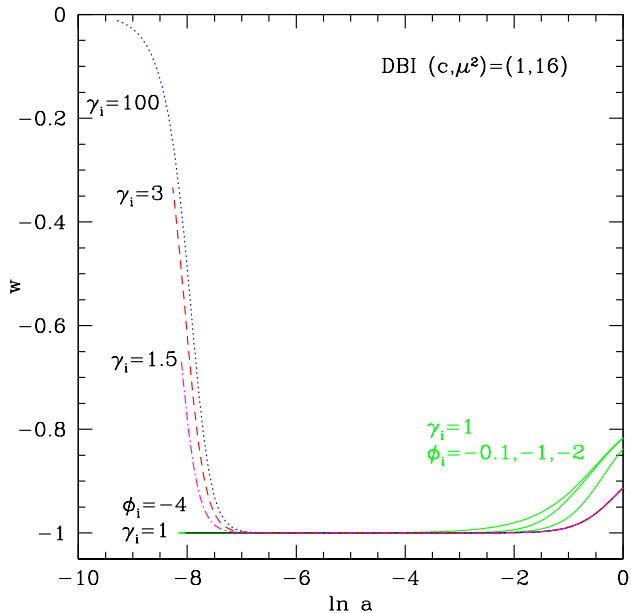,width=3.4in}
\caption{The high Hubble friction during matter domination freezes the 
field to $w=-1$ for many e-folds in expansion, despite an initial field 
velocity measured in terms of the Lorentz factor $\gamma_i$.  This 
pseudo-attractor ensures that models with different $\gam_i$ then follow 
the same trajectory, as shown by the convergence of tracks from the 
left side (early times) to the middle (later times).  (Tracks start 
in the plot at $\Omega_\phi=10^{-10}$, with $\phi_i=-4$ and the 
$\gamma_i$ as labeled.)  The light, 
green curves diverging from the middle to the right side (today) show 
this is not a true attractor since the thawing rate does 
depend on the initial field value $\phi_i$ (here fixing $\gamma_i=1$).  
However, the late time, 
true attractor from DBI dynamics will bring all these trajectories 
together; indeed at present all models with $\kappa\phi_i<1$ have the 
same dynamics. 
}
\label{fig:gami}
\end{center}
\end{figure}

At early times, in 
the matter dominated era, the field is frozen to a cosmological constant 
state, until the DBI dark energy density become nonnegligible.  This 
is independent of initial field value $\phi_i$ and velocity $\gam_i$, 
as Fig.~\ref{fig:gami} 
illustrates.  The freezing represents the effects of matter domination 
and is a different sort of attractor than the late time solution. 
The thawing occurs in a manner that does depend on $\phi_i$, but is 
insensitive today to $\phi_i$ for $|\phi_i|<1$.  In the future, the 
DBI attractor ensures the same solution regardless of $\phi_i$.

At late times the field only notices the quadratic part of the 
potential; that is, this attractor solution only requires that the 
potential look quadratic near the minimum -- a highly generic state. 
The evolution of the field up to the present, however, does depend on 
the quartic term: contrast the $(c,\mu^2)=(0,16)$ and $(1,16)$ curves 
in Fig.~\ref{fig:cdmodels}.  
At all times until the final asymptotic 
value the specific evolution differs, in particular up to the present. 

These differences allow us to constrain the parameters of the theory 
by comparing to cosmological observations.  Here we consider the 
distance-redshift relation over the range $z=0-1.7$, as given by 
Type Ia supernovae.  We examine the maximum fractional difference 
$\delta d/d_\Lambda$ of the model predictions for distances from those 
of the flat, cosmological constant plus matter universe with 
$\Omega_\Lambda=1-\Omega_m=0.72$.

One question we can ask is what are the bounds on $\mu^2$ such that 
the distance deviation is less than some value, say 2\%.  Large values 
of $\phi_i$ give a lengthy frozen state (note $(1/V)[dV/d(\kappa\phi)] 
\sim 1/\phi$ becomes small), lasting until close to the present, so 
$w\approx-1$.  This will give little deviation from a cosmological 
constant so the most stringent bounds on $\mu^2$ occur for small $\phi_i$.  
For $\kappa\phi_i\lesssim1$, though, the potential tends to be dominated 
by the quadratic, attractor part and the field quickly forgets the 
initial value (compare the $\kappa\phi_i=-0.1$ vs.\ $\kappa\phi_i=-1$ 
curves in Fig.~\ref{fig:gami}).  This also makes the bound fairly 
insensitive to the value of $c$.  A fitting formula to the constraint on 
$\mu$ is 
\beq 
\mu>7.1\,(1+0.002c)\left(\frac{\delta d/d_\Lambda}{2\%}\right)^{-1}\,. \label{eq:mubound} 
\eeq 
Note the weak dependence on $c$.  The inverse proportionality to 
$\delta d/d_\Lambda$, for small deviations, arises from the maximum 
deviation in the equation of state $1+w$.  The attractor value is 
given by \cite{akl} 
\beq 
1+w_c=\frac{2}{3\mu^2}\left[-1+\sqrt{1+3\mu^2}\right]\,, 
\eeq 
which is inversely proportional to $\mu$, for $\mu^2\gg1$. 

\begin{figure}[!htb]
\begin{center}
\psfig{file=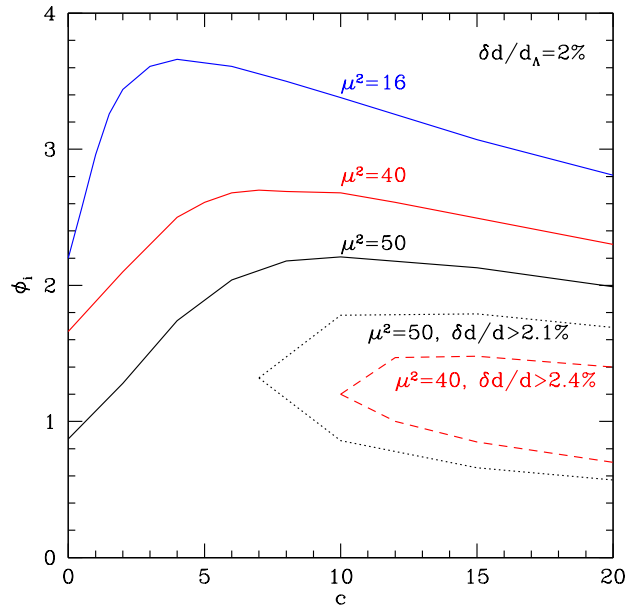,width=3.4in}
\caption{Model parameters can be constrained by comparison to distance 
data, here taken to be within 2\% of $\Lambda$CDM.  Above the solid curves 
for each value of $\mu^2$ the deviation is less than 2\% (as $\phi_i$ 
gets large, $w$ has deviated less from the value $w=-1$ imposed by 
the matter dominated freezing).  Below the solid curves the distance 
deviation is larger, but often by a small amount: only within the 
dotted, black contour is the deviation more than 2.1\% for $\mu^2=50$, 
and similarly the dashed, red contour bounds the deviation to 2.4\% for 
$\mu^2=40$. 
}
\label{fig:muphic}
\end{center}
\end{figure}

While Eq.~(\ref{eq:mubound}) gives the most stringent bound to agree 
with observations, models with lesser values of $\mu$ are viable if 
the values of $\phi_i$ are large enough.  Figure~\ref{fig:muphic} 
show the constraints in the $c$-$\phi_i$ plane for a maximum allowed 
distance deviation of 2\%.  For $\mu^2\gtrsim55$, the distance deviation 
is less than 2\% for all cases with $c<20$.  The maximum tends to be 
quite shallow: for $\mu^2=50$ most of the disallowed lower half plane 
actually has $0.02<\delta d/d<0.021$.  The largest deviation for 
$\mu^2=50$ (40) occurs for $c=20$ and is at the 2.18\% (2.44\%) level. 
The figure exemplifies how cosmological observations can directly 
inform us on string theory parameters. 

\section{Customized Expansion History \label{sec:exphis}} 

From Eq.~(\ref{eq:wdef}), we can write down a solution for the form 
of the potential for any expansion history desired, i.e.\ any given 
equation of state evolution $w(a)$ (including $w$ constant). 
The reduced potential, $v=V(\phi)/T(\phi)$ must satisfy 
\beq
v(a)=1-\frac{w(a)\,\gamma(a)}{1+w(a)}-\frac{\gamma(a)^{-1}}{1+w(a)} 
\,. \label{eq:vwcon}
\eeq
Note this expression holds even for a time dependent $\gamma$ (we are 
here interested in the full evolution, not just the attractor state). 
Combining this expression for $v(a)$ with the solutions of the equations 
of motion for $\gamma(a)$ and $\phi(a)$, one can construct the potential 
$V(\phi)$ for any desired equation of state function. 

\begin{figure}[!htb]
\begin{center}
\psfig{file=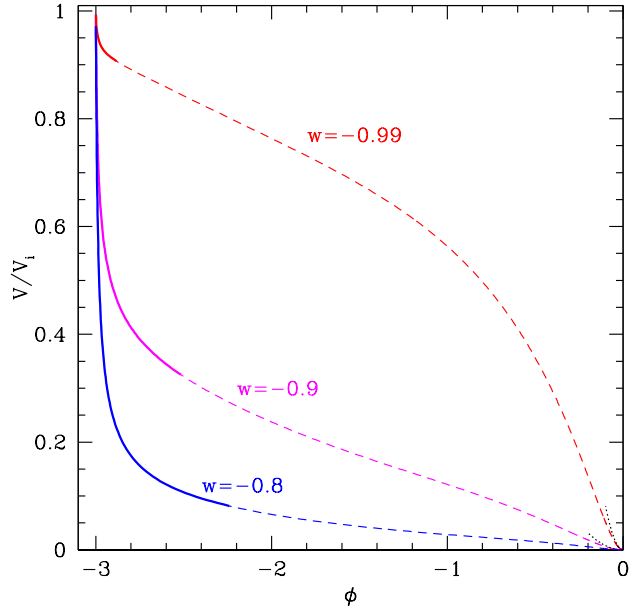,width=3.4in}
\caption{Solutions for the potential function are exhibited that deliver 
constant values of the equation of state $w$.  Solid portions of the curves 
correspond to the region over which the field has rolled by the 
present.  Short dotted arcs near $\phi=0$ show the $V\sim\phi^2$ 
asymptotic behavior.  The potentials do not contain an 
explicit cosmological constant (i.e.\ $V(\phi)$ has a true zero minimum), 
but the equations of state can approach $w=-1$ due to the DBI dynamics. 
} 
\label{fig:wconw}
\end{center}
\end{figure}

Figure~\ref{fig:wconw} shows the potential $V(\phi)$ constructed 
(taking $T\sim\phi^4$) to 
give constant $w$ for all times, for the cases $w=-0.99$, $-0.9$, and 
$-0.8$.  (If $w=-1$ exactly then the field does not roll at all and 
the potential cannot be reconstructed.)  
The conditions for $w\approx-1$ to be realized (for constant $w$) 
can be written through Eq.~(\ref{eq:wdef}) in terms of the initial 
values (note we are not describing an attractor solution) and are that either 
$v_i\gg \gamma_i$ or $v_i\gg \gamma_i-1$.  For $\gamma_i=1+\epsilon$, 
with $\epsilon$ a small quantity, 
$w\approx -1+2\epsilon/v_i$ if the second condition holds.  When the 
first condition holds, $w\approx -1+[(\gamma_i^2-1)/\gamma_i](1/v_i)$. 
In either case, $w\to-1$.  The potential is steep initially (roughly 
$\lam^2\sim\Omega_{\phi,i}^{-1}$) and the field 
rolls to $\phi=0$.  The shape of the potential near $\phi=0$ is given 
by $V(\phi\ll1)\sim\phi^2$ (since we took $T\sim\phi^4$), as required 
by our previous results.  However, as noted there, the potential when 
the dynamics is off the attractor trajectory does not need to stay in 
the asymptotic form.

\section{Multi-Brane DBI \label{sec:wdbi}}

In the presence of multiple D3-branes or a non-BPS brane, the DBI 
action acquires an additional potential $U(\phi)$ multiplying the DBI term 
\cite{gumward,saridakis}, 
\beqa 
S&=&\int d^4x\,\sqrt{-g} \,\times \nonumber \\ 
&\,&\left[-U(\phi)\,T(\phi)\sqrt{1-\dot\phi^2/T(\phi)}+T(\phi)-V(\phi)\right]. \label{eq:lagu} 
\eeqa 
The energy-momentum tensor
takes a perfect fluid form with energy density $\rho_\phi$ and pressure
$p_\phi$ given by
\beq
\rho_\phi=(\gamma U-1)\,T+V \quad ;\quad p_\phi=(1-\gamma^{-1}U)T-V\,. \label{eq:rhodef} 
\eeq
The Lorentz factor $\gamma$ is still given by Eq.~(\ref{eq:gamdef}) and 
the equation of state for the DBI field is
\beq
w\equiv \frac{p_\phi}{\rho_\phi}=-\frac{\gamma^{-1}U-1+v}{\gamma U-1+v}
\,. \label{eq:wudef}
\eeq
The extra freedom from the additional potential $U$ means that 
interesting results occur in both the nonrelativistic and relativistic 
cases, not just $\gamma\to\infty$ as in the standard DBI model.

\subsection{Equations of Motion and Attractors \label{sec:wattrx}} 

The equation of motion for the field follows from either functional
variation of the action or directly from the continuity equation for
the energy density,
\beq
\rho'_\phi=-3(\rho_\phi+p_\phi)=-3(\gamma -\gamma^{-1})\,UT\,,
\eeq
where a prime denotes a derivative with respect to the e-folding parameter,
$d/d\ln a$.  

To begin, we define the contributions of the tension and potential 
to the vacuum energy density relative to the critical density, 
\beq 
x^2=\frac{\kappa^2}{3H^2}(\gamma U-1)\,T \quad ; \quad
y^2=\frac{\kappa^2}{3H^2}V \,, \label{eq:xdef}
\eeq 
where $\kappa^2=8\pi G$ and $H$ is the Hubble parameter.  We allow 
the parameter $x^2<0$ so as to unify the treatment of when $\gam U>1$ 
and $\gam U<1$. 
The equations of motion are given by 
\beqa 
\frac{1}{2}(x^2)'&=&-\frac{3}{2}x^2(1-x^2)\frac{1-\gamma^{-1}U}{\gamma U-1}
-\frac{3}{2}x^2y^2 \nonumber \\ 
&\,&+\frac{\sqrt{3}}{2}\lambda\,y^2\sqrt{\frac{(\gamma^2-1)\,x^2}{\gamma^2(\gamma U-1)}}\\
y'&=&\frac{3}{2}\,x^2y\,\frac{\gamma^2-1}{\gamma}\frac{U}{\gamma U-1}
+\frac{3}{2}\,y\,(1-x^2-y^2) \nonumber \\ 
&\,&-\frac{\sqrt{3}}{2}\lambda\,y\sqrt{\frac{(\gamma^2-1)\,x^2}{\gamma^2(\gamma U-1)}} \\ 
\kappa\phi'&=&\sqrt{\frac{3(\gamma^2-1)\,x^2}{\gamma^2(\gamma U-1)}}\,, 
\eeqa 
where $\lam=-(1/\kappa V)dV/d\phi$ and 
\beq 
\gamma U=1+\frac{V}{T}\frac{x^2}{y^2}\,. \label{eq:gamv} 
\eeq 
When $U=1$ these equations reduce to those in \cite{akl}.  The case 
$\gam U=1$ can be handled by the above equations since the denominator 
$\gam U-1$ always occurs in the finite ratio $x^2/(\gam U-1)$. 

We are interested in the DBI field as late time accelerating dark
energy, not for inflation, so we take the initial conditions in the
matter dominated universe and define the present by
$\Omega_\phi=0.72$. 
The attractor solutions to the equations of motion have the critical values  
\beqa 
x^2_{c1}&=&\frac{\lam^2}{3U^2}\frac{\gamma U-1}{\gamma^2-1} \quad ; \quad 
x^2_{c2}=\frac{3\gamma^2}{\lam^2}\frac{\gam U-1}{\gam^2-1} \\ 
y^2_{c1}&=&1-\frac{\lam^2}{3U^2}\frac{\gam U-1}{\gam^2-1} \quad ; \quad 
y^2_{c2}=\frac{3}{\lam^2}\frac{\gam^2-\gam U}{\gam^2-1} \\ 
\Omega_{\phi,c1}&=&1 \quad ; \quad \Omega_{\phi,c2}=\frac{3\gam U}{\lam^2} \label{eq:ophic} \\ 
w_{\phi,c1}&=&-1+\frac{\lam^2}{3\gam U} \quad ; \quad w_{\phi,c2}=0 \,. \label{eq:wc} 
\eeqa 
These are stable, late time attractors, with the $w\ne0$ solution reached 
for $\lambda^2<3\gamma U$.  The form of these solutions 
reveals that paths to the attractor classes are more diverse compared 
to standard DBI theory.  For example, new windows appear for obtaining 
$w=-1$ if $U(\phi_c)\to\infty$ sufficiently quickly.  In particular, 
this cosmological constant behavior can even be realized when $\gam\to1$, 
without the potential running to infinite field values.  
Now the important limit is when $\gam U\to\infty$ rather than 
$\gam$ alone.  These attractors can therefore be achieved when $\gam$ 
remains nonrelativistic but $U$ gets large for the asymptotic field value. 

The attractor value for $w$ depends on two key parameters: $\lam^2/U$ 
and $v\lam^2/U^2$.  The explicit solution is given by 
\beq 
w=-1+2\,\left[1+\sqrt{1+12\,\frac{v-1}{\lam^2}+\left(\frac{6U}{\lam^2}\right)^2}\right]^{-1}\,, \label{eq:wulim} 
\eeq 
and the value of the Lorentz boost factor is 
\beq 
\gam=\frac{\lam^2}{6U}+\sqrt{\left(\frac{\lam^2}{6U}\right)^2 
+\frac{\lam^2\,(v-1)}{3U^2}+1}\,. \label{eq:gamlim} 
\eeq 

Table~\ref{tab:ucrit} shows the parameter combinations that lead to 
attractors with accelerated expansion.  As stated, although the 
essential classes of attractors (the four groups divided by the 
horizontal rows)  are the same as with standard DBI (cf.~Table~1 
of \cite{akl}), the {\it paths\/} to obtaining them are multiplied. 
These can deliver cosmological constant like behavior nonrelativistically, due 
to the influence of the multibrane potential $U$, as well as new 
approaches to $w=$ constant, arbitrarily close to $w=-1$. 
(However, as we discuss in the next subsection, one can also absorb 
$U$ into standard DBI.)

\begin{table}[htbp]
\begin{center}
\begin{tabular*}{0.9\columnwidth}
{@{\extracolsep{\fill}} c c c c c c}
\hline
$V/T$ & $\lam^2/U$ & $\lam^2v/U^2$ & $\gam$ & $\gam U$ & $w$ \\ 
\hline
$\infty$ & moot & $\infty$ & $\infty$ & $\infty$ & $-1$ \\
$\infty$ & $0$ & 0 & 1 & $\infty$ & $-1$ \\ 
\hline 
$\infty$ & $\infty$ & $\infty$ & $\infty$ & $\infty$ & const \\ 
$\infty$ & $\infty$ & const & $\infty$ & const & const \\ 
$\infty$ & const & const & const & const & const \\ 
$\infty^\dagger$ & 0 & 0 & 1 & $\infty$ & $-1$ \\ 
\hline 
const & const & const & const & const & const \\ 
const & $\infty$ & $\infty$ & $\infty$ & const & const \\ 
const & 0 & 0 & 1 & $\infty$ & $-1$ \\ 
\hline 
0 & const & 0 & 1 & const & const \\ 
0 & 0 & 0 & $1$ & const${}^*$ & $-1$ \\ 
\hline 
\end{tabular*}
\end{center}
\caption{Summary of accelerating attractor properties.  The columns
give the values of the quantities for the attractor solution, all of 
which possess asymptotic $\Omega_\phi=1$.  Each grouping of rows corresponds 
to one of the classes of standard DBI from Table~1 of \cite{akl}, with 
the first row of each group being the standard DBI solution.  We see 
that multibrane DBI increases the number of ways of obtaining 
accelerating attractor 
solutions by almost a factor 3 over standard DBI and a factor 11 over 
quintessence.  The dagger indicates that while $V/T=\infty$, $(V/T)/\lam^2=$ 
const.  The asterisk in the last row denotes that the constant 
is 0 unless $U\to\infty$.  The values of constant $w$ are given by 
Eq.~(\ref{eq:wulim}).}  
\label{tab:ucrit}
\end{table}

Class 1 in the first group of rows of the table achieves cosmological 
constant behavior. This can be realized, for example, through taking 
$T\sim\phi^m$, 
$V\sim\phi^c$, $U\sim\phi^p$ with any $p<-2$.  In other words, even 
forms of the tension $T$ and potential $V$ that in standard DBI do 
not give acceleration, let alone $w=-1$, can give an asymptotic 
cosmological constant state if $U$ increases sufficiently rapidly, e.g.~having 
an inverse power law form with $p<-2$.  The steepness of $U$ trumps the 
behavior of $V$, $T$ so also the standard case giving $w$ constant 
(e.g.~$T\sim\phi^4$, $V\sim\phi^2$) would instead yield $w=-1$. 

Class 2 in the second group of rows of the table delivers a constant $w$, 
which can be made arbitrarily close to $-1$ depending on parameter values. 
An example would be given by the additional multibrane potential with $p=-2$. 
Here, though, if $V$ and $T$ were such that they would cause an attractor 
to $w=-1$, then this still holds.  Alternately, if $V$ and $T$ could not 
attain an accelerating attractor, $U\sim\phi^{-2}$ can achieve this with 
a constant $w$.  Note that the presence of $U$ also alters the value of 
constant $w$ (cf.\ Eq.~\ref{eq:wulim}) from the standard DBI case where 
$V$, $T$ give a constant $w$. 

\begin{figure}[!htb]
\begin{center}
\psfig{file=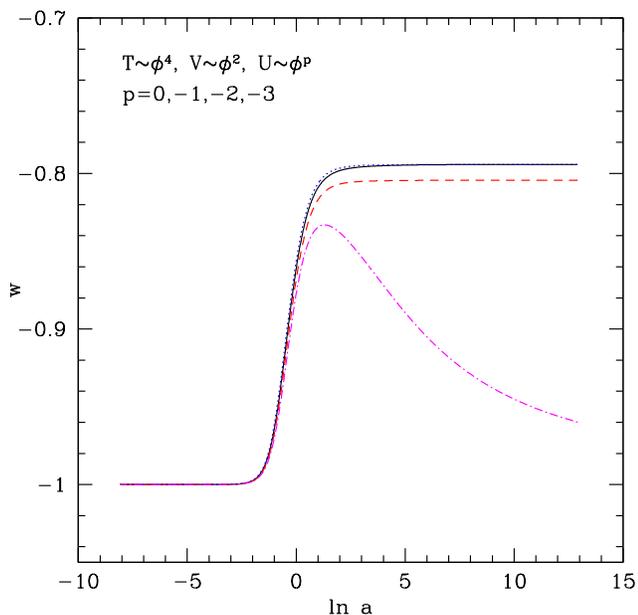,width=3.4in}
\caption{The presence of the multibrane potential $U$ alters the 
conditions for attractor solutions and opens up new routes to approach $w=-1$. 
When $\lam^2/U\to0$, then the cosmological constant is the 
asymptotic solution.  When this combination goes to a constant value, 
then $w\to$ constant given by Eq.~(\ref{eq:wulim}), and when the 
combination goes to $\infty$ then the standard DBI solution is unchanged.  
For a power law 
potential $U\sim\phi^p$, these correspond to $p<-2$, $=-2$, $>-2$. 
}
\label{fig:att42p}
\end{center}
\end{figure}

\begin{figure}[!htb]
\begin{center}
\psfig{file=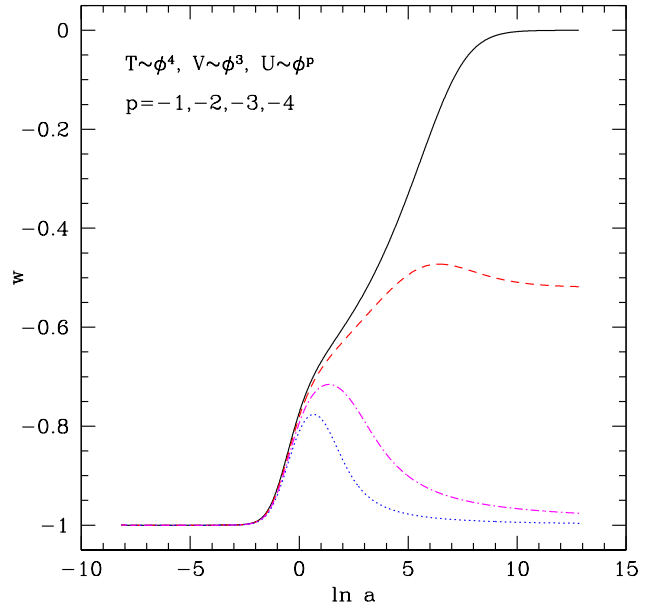,width=3.4in}
\caption{As Fig.~\ref{fig:att42p} but for a form of the standard 
potential $V$ that would not give an accelerating attractor in 
standard DBI theory.  Here, however, the multibrane potential can 
give constant $w$ (for $p=-2$) or a cosmological constant (for $p<-2$). 
}
\label{fig:att43p}
\end{center}
\end{figure}

However, if $U$ does not get large sufficiently quickly, e.g.~$p>-2$, 
then $V$ and $T$ determine the attractor behavior in the same manner 
as in standard DBI.  Figures~\ref{fig:att42p} and \ref{fig:att43p} 
illustrates these various behaviors, for cases where standard DBI would 
predict a constant $w$ attractor and no accelerating attractor, 
respectively.  (We do not show the $V\sim\phi^1$ case because as stated 
this has identical asymptotic behavior to the standard DBI theory.)

Class 3 is characteristic of exponential potential and tension, where 
the field runs off to infinity.  However, the behavior of $U$ can 
determine the value of $w$, leading to either a constant $w\ne-1$ 
attractor or 
a cosmological constant state, unlike in standard DBI.  Class 4 is similar 
to standard quintessence but again $U$ can deliver $w=-1$. 

Just as in \S\ref{sec:exphis}, one can design a function $U$ to fit a 
specific expansion history, or equation of state, through 
Eq.~(\ref{eq:wudef}).  Also note that 
a constraint on $U$ exists from the nonnegativity of the 
energy density in Eq.~(\ref{eq:rhodef}).  This imposes the condition 
\beq 
\gam U \ge 1-v\,. \label{eq:ucond}
\eeq 
This is automatically satisfied for $\gam U\ge1$ 
(we always take $V$, $T$ nonnegative).  For $\gam U<1$ though it limits 
the allowed forms of $U(\phi)$. 
When $\gam U=1$ then $w=-1+\lam^2/3$ at all times, not just asymptotically 
(when $\lam^2>3$ there is no attractor).  This looks like a standard 
quintessence attractor solution, but can actually be realized by 
a relativistic $\gam$ model with $U<1$.

\subsection{Single Brane Equivalence \label{sec:nonrel}} 

In examining the nonrelativistic limit of the action (\ref{eq:lagu}) 
we see that it approaches quintessence with a redefinition of the 
field and potential.  This suggests a deeper mapping between the 
multibrane and standard single brane DBI actions.  By defining 
\begin{equation} \label{varphi}
\varphi \equiv \int \sqrt{U} d\phi
\end{equation}
we can rewrite the action \eqref{eq:lagu} in terms of $\varphi$: 
\begin{equation}\label{eq:lagu2} 
S=\int d^4x\,\sqrt{-g} \left[-TU\sqrt{1-\dot\varphi^2/(TU)}+T-V\right]. 
\end{equation}
Comparing this action with Eq.~\eqref{eq:lag}, we see that it
is equivalent to the original DBI action with tension $\hat T$ and potential 
$\hat V$ given by
\begin{align}
\hat T &= TU,  \\
\hat V &= TU - T + V\,.
\end{align}
Therefore, the general analysis of \cite{akl} applies to the multibrane 
situation when viewed in terms of the equivalent single brane, hatted 
quantities. Specifically, 
the formulae \eqref{eq:xdef}--\eqref{eq:wc} hold for
\beq 
x^2=\frac{\kappa^2}{3H^2}(\gamma -1)\,\hat T \quad ; \quad
y^2=\frac{\kappa^2}{3H^2}\hat V \,, \label{eq:xdefh}
\eeq 
with the replacement $U \rightarrow 1$, $v\to\hat v$, $\phi\to\varphi$, 
and $\lam\to\lamh$ where $\lamh=-(1/\hat V)d\hat V/d(\kappa\varphi)$.  
In this formulation, the attractor values for $w$ and $\gamma$, 
Eqs.~(\ref{eq:wulim}) and (\ref{eq:gamlim}), take 
the same form as in standard DBI. 

As an explicit example of the mapping between the multibrane and 
single brane views, let us consider the case where the (unhatted) 
tension and the potentials are given by power laws,
\begin{equation}
T \sim \phi^m,\quad V \sim \phi^c,\quad U \sim \phi^p,
\end{equation}
and investigate how the attractor values of $\gamma$ and $w$ change
as the exponents are varied.  This gives an alternate view and 
derivation of the results in Sec.~\ref{sec:wattrx}.  
We assume $m$ and $c$ are positive for
simplicity. From Eq.~\eqref{varphi}, the redefined field $\varphi$
is related to the original field $\phi$ as $\varphi \sim \phi^{(p+2)/2}$ 
and the hatted quantities become
\begin{align}
\hat T &= TU \sim \varphi^{\frac{2(m+p)}{p+2}}, \notag \\
\hat V &= TU - T + V 
       \sim \varphi^{\frac{2(m+p)}{p+2}}
            - \varphi^{\frac{2m}{p+2}} + \varphi^{\frac{2c}{p+2}}, \notag \\
\hat v &= \hat V/ \hat T
       \sim 1 - \varphi^{-\frac{2p}{p+2}} + \varphi^{\frac{2(c-m-p)}{p+2}}.
\end{align}
Note that, if $p<-2$, $\varphi$ is inversely proportional to $\phi$ and
the small-field limit for one is the large-field limit for the other. 
Thus it is natural to separately study the cases $p>-2$ and $p<-2$. 

For the case $p>-2$, all the powers of the terms in $\hat V$ are positive
and $\varphi$ would go to zero asymptotically. Then the logarithmic derivative
$\lamh \sim 1/\varphi$ diverges, giving the ultrarelativistic class of
attractor solution $\gamma \rightarrow \infty$. To obtain $w=-1$, 
$\hat v/\lamh^2$ should diverge, which happens if $m-c>2$. Note that
this result is independent of $p$. Therefore we conclude that
if $U$ is less singular than $1/\phi^2$ there is no effect of $U$, 
in agreement with Sec.~\ref{sec:wattrx}. 

If $p=-2$, then $\phi\sim e^\varphi$ and the hatted potentials and tension 
appear exponential.  These give constant $w$ attractors, even if (unhatted) 
$V$ and $T$ would not normally give acceleration.  If $V$ and $T$ would 
give $w=-1$ by themselves, then this is maintained. 

If $p<-2$, then as noted above the small-field and large-field limits 
are reversed.  Thus we obtain $w=-1$ in any case: if $V$ and $T$ provide 
$w=-1$ themselves, then this is maintained, while if they do not give 
acceleration then $U$ operates in the opposite limit and drives the 
field to a $w=-1$ attractor.  Again, see 
Figs.~\ref{fig:att42p}-\ref{fig:att43p} and Sec.~\ref{sec:wattrx}. 

As a curiosity, note we could take the converse view and split the 
single brane picture into multiple branes.  For example, the usual 
quartic single brane tension $\hat T\sim\varphi^4$ could be viewed 
as $T\sim\phi^m$ and $U\sim\phi^{m-4}$ as a way of relaxing the conditions 
on the brane tension.  It is this extra freedom from $U$ that generates 
further paths to the same attractors as in standard DBI. 

Another interesting case arises by choosing $T=V=\text{const}$ and
$U(\phi)$ as a runaway type potential connecting $U(0)=1$ and $U(\infty)=0$.
Then the action can be interpreted as the action for an unstable D-brane in 
string theory \cite{sen} and the field $\phi$ represents its tachyonic mode.
A standard form for $U(\phi)$ is \cite{buchel,kims,kutasov} 
\begin{equation}
U = 1/\cosh{\alpha\phi}
\end{equation}
where $\alpha$ is a constant. In this case, $\varphi \sim e^{-\alpha\phi/2}$
and $\hat V \sim \varphi^2$. Then we get $\gamma \to \infty$ and 
$w =0$.

\section{Sound Speed \label{sec:sound}} 

Beyond the homogeneous field properties we can briefly consider perturbations 
to the dark energy density.  These propagate with sound speed $c_s$ and 
define a Jeans wavelength above which the dark energy can cluster.  The 
sound speed is defined in terms of the Lagrangian density $L$ (given by 
the term in square brackets in Eq.~\ref{eq:lag} or \ref{eq:lagu}) and 
canonical kinetic energy $X=(1/2)\dot\phi^2$ as \cite{soundspeed} 
\beq 
c_s^2=\frac{L_{,X}}{L_{,X}+2XL_{,XX}}\,, 
\eeq 
The result is $c_s=1/\gam$ for both the standard \cite{martinyam} and 
generalized DBI 
actions, since $U(\phi)$ does not change the kinetic structure. 

For the attractors depending on the relativistic limit, such as for 
$w\approx-1$ in the standard DBI case, this implies the sound speed 
goes to 0 and dark energy can clump on all scales.  One of the 
interesting aspects of multibrane DBI is that this is no longer 
necessary; $w=-1$ can be achieved with $\gamma=1$ and so $c_s=1$. 
However, when $w\approx-1$ in whichever case then dark energy 
perturbations cannot grow regardless of the sound speed, so the 
sound speed is unlikely to give a clear signature of the 
DBI theory for the cases we consider.  Indeed even models of dark 
energy with $c_s=0$ cannot 
be readily distinguished from those with $c_s=1$, when $w\approx-1$ 
and the dark energy does not couple to matter 
\cite{beandore,huscranton,coragm} 
(see \cite{matarrese,lazkoz} for the case of coupling).

\section{Conclusions \label{sec:concl}} 

We have investigated possible constraints on DBI string theory from 
cosmological observations, considering the entire field evolution not 
just the asymptotic future behavior.  
In particular, Eq.~(\ref{eq:mubound}) gives 
a bound on the deviation of the locally warped region generated by the 
form-field fluxes from the AdS geometry.  It is very interesting if more 
accurate cosmological data can restrict fundamental string parameters. 

To improve the fine tuning problem of initial conditions, we have 
enlarged attractor solutions to the case of generalized DBI theory which 
includes an additional potential arising from either multiple coincident 
branes, or non-BPS branes, or D5-branes wrapping a two-cycle within the 
compact space and carrying a non-zero magnetic flux along this cycle 
\cite{gumward}.  We have obtained exact cosmological constant behavior 
from some attractors of the extended DBI theory.  Also, we have noticed 
that the extended DBI theory can have the identical attractor behavior 
to single-brane DBI with a different tension and potential.  

An interesting novel feature of the DBI attractors is that the 
sound speed can be driven to zero which enhances dark energy 
clustering, although this is suppressed when $w\approx-1$. 
We also showed that a straightforward quadratic plus quartic potential 
acts like a thawing scalar field, and how more complicated potentials 
could be designed for a specific cosmic expansion history. 

We have analyzed in greater detail than in \cite{akl} how accurate 
cosmological observations on the dark energy can constrain some 
aspects of fundamental string theory within the DBI framework.  
Input from high energy physics on the forms of the functions 
is necessary as well. 
The connections between string theory and astrophysical data offer 
exciting prospects for revealing the nature of the cosmological constant 
and the accelerating universe.

\acknowledgments

This work has been supported by the World Class University grant
R32-2008-000-10130-0. CK has been supported in part by the KOSEF grant
through CQUeST with grant No.\ R11 - 2005 - 021. 
EL has been supported in part by the
Director, Office of Science, Office of High Energy Physics, of the
U.S.\ Department of Energy under Contract No.\ DE-AC02-05CH11231.

\end{document}